\documentstyle[12pt]{article}

\def\hybrid{\topmargin -20pt    \oddsidemargin 0pt
        \headheight 0pt \headsep 0pt
        \textwidth 6.25in       
        \textheight 9.5in       
        \marginparwidth .875in
        \parskip 5pt plus 1pt   \jot = 1.5ex}

\hybrid

\def\baselinestretch{1.2}

\catcode`\@=11

\def\marginnote#1{}
\newcount\hour
\newcount\minute
\newtoks\amorpm
\hour=\time\divide\hour by60
\minute=\time{\multiply\hour by60 \global\advance\minute by-\hour}
\edef\standardtime{{\ifnum\hour<12 \global\amorpm={am}%
        \else\global\amorpm={pm}\advance\hour by-12 \fi
        \ifnum\hour=0 \hour=12 \fi
        \number\hour:\ifnum\minute<10 0\fi\number\minute\the\amorpm}}
\edef\militarytime{\number\hour:\ifnum\minute<10 0\fi\number\minute}

\def\draftlabel#1{{\@bsphack\if@filesw {\let\thepage\relax
   \xdef\@gtempa{\write\@auxout{\string
      \newlabel{#1}{{\@currentlabel}{\thepage}}}}}\@gtempa
   \if@nobreak \ifvmode\nobreak\fi\fi\fi\@esphack}
        \gdef\@eqnlabel{#1}}
\def\@eqnlabel{}
\def\@vacuum{}
\def\draftmarginnote#1{\marginpar{\raggedright\scriptsize\tt#1}}

\def\draft{\oddsidemargin -.5truein
        \def\@oddfoot{\sl preliminary draft \hfil
        \rm\thepage\hfil\sl\today\quad\militarytime}
        \let\@evenfoot\@oddfoot \overfullrule 3pt
        \let\label=\draftlabel
        \let\marginnote=\draftmarginnote
   \def\@eqnnum{(\theequation)\rlap{\kern\marginparsep\tt\@eqnlabel}%
\global\let\@eqnlabel\@vacuum}  }

\def\preprint{\twocolumn\sloppy\flushbottom\parindent 2em
        \leftmargini 2em\leftmarginv .5em\leftmarginvi .5em
        \oddsidemargin -.5in    \evensidemargin -.5in
        \columnsep .4in \footheight 0pt
        \textwidth 10.in        \topmargin  -.4in
        \headheight 12pt \topskip .4in
        \textheight 6.9in \footskip 0pt
        \def\@oddhead{\thepage\hfil\addtocounter{page}{1}\thepage}
        \let\@evenhead\@oddhead \def\@oddfoot{} \def\@evenfoot{} }

\def\numberbysection{\@addtoreset{equation}{section}
        \def\theequation{\thesection.\arabic{equation}}}

\def\underline#1{\relax\ifmmode\@@underline#1\else
        $\@@underline{\hbox{#1}}$\relax\fi}

\def\titlepage{\@restonecolfalse\if@twocolumn\@restonecoltrue\onecolumn
     \else \newpage \fi \thispagestyle{empty}\c@page\z@
        \def\thefootnote{\fnsymbol{footnote}} }

\def\endtitlepage{\if@restonecol\twocolumn \else \newpage \fi
        \def\thefootnote{\arabic{footnote}}
        \setcounter{footnote}{0}}  

\catcode`@=12
\relax

\def\figcap{\section*{Figure Captions\markboth
        {FIGURECAPTIONS}{FIGURECAPTIONS}}\list
        {Figure \arabic{enumi}:\hfill}{\settowidth\labelwidth{Figure
999:}
        \leftmargin\labelwidth
        \advance\leftmargin\labelsep\usecounter{enumi}}}
 \relax
\def\tablecap{\section*{Table Captions\markboth
        {TABLECAPTIONS}{TABLECAPTIONS}}\list
        {Table \arabic{enumi}:\hfill}{\settowidth\labelwidth{Table
999:}
        \leftmargin\labelwidth
        \advance\leftmargin\labelsep\usecounter{enumi}}}
 \relax
\def\reflist{\section*{References\markboth
        {REFLIST}{REFLIST}}\list
        {[\arabic{enumi}]\hfill}{\settowidth\labelwidth{[999]}
        \leftmargin\labelwidth
        \advance\leftmargin\labelsep\usecounter{enumi}}}
 \relax
\makeatletter
\newcounter{pubctr}
\def\publist{\@ifnextchar[{\@publist}{\@@publist}}
\def\@publist[#1]{\list
        {[\arabic{pubctr}]\hfill}{\settowidth\labelwidth{[999]}
        \leftmargin\labelwidth
        \advance\leftmargin\labelsep
        \@nmbrlisttrue\def\@listctr{pubctr}
        \setcounter{pubctr}{#1}\addtocounter{pubctr}{-1}}}
\def\@@publist{\list
        {[\arabic{pubctr}]\hfill}{\settowidth\labelwidth{[999]}
        \leftmargin\labelwidth
        \advance\leftmargin\labelsep
        \@nmbrlisttrue\def\@listctr{pubctr}}}
 \relax
\makeatother
\newskip\humongous \humongous=0pt plus 1000pt minus 1000pt

\newif\ifdtup

\relax

\def\be{\begin{equation}}
\def\ee{\end{equation}}
\def\ba{\begin{eqnarray}}
\def\ea{\end{eqnarray}}

\begin{document}

\renewcommand{\theequation}{\arabic{equation}}

\newcommand{\beq}{\begin{equation}}
\newcommand{\eeq}[1]{\label{#1}\end{equation}}
\newcommand{\ber}{\begin{eqnarray}}
\newcommand{\eer}[1]{\label{#1}\end{eqnarray}}
\newcommand{\eqn}[1]{(\ref{#1})}
\begin{titlepage}
\begin{center}

\hfill gr--qc/0504130\\
\vskip -.1 cm
\hfill CERN-PH-TH/2005-069\\
\vskip -.1 cm
\hfill April 2005\\

\vskip .8in

{\large \bf On the integrability of spherical gravitational waves in vacuum} 

\vskip 0.8in

{\bf Ioannis Bakas}\footnote{On sabbatical leave from   
Department of Physics, University of Patras, GR-26500 Patras, Greece; e-mail: 
bakas@ajax.physics.upatras.gr}
\vskip 0.1in
{\em Theory Division, Department of Physics, CERN\\
CH-1211 Geneva 23, Switzerland\\
\footnotesize{\tt ioannis.bakas@cern.ch}}\\

\end{center}

\vskip 1in

\centerline{\bf Abstract}

\noindent
The general class of Robinson-Trautman metrics that describe gravitational 
radiation in the exterior of bounded sources in four space-time dimensions
is shown to admit zero curvature formulation in terms of appropriately 
chosen two-dimensional gauge connections. The result, which is valid  
for either type II or III metrics, implies that the gravitational 
analogue of the Lienard-Wiechert fields of Maxwell equations form a new  
integrable sector of Einstein equations for any value of the 
cosmological constant. The method of investigation is 
similar to that used for integrating the Ricci flow in two dimensions. 
The zero modes of the gauge symmetry (factored by the center) generate 
Kac's $K_2$ simple Lie algebra with infinite growth.    
\vfill
\end{titlepage}
\eject

\def\baselinestretch{1.2}
\baselineskip 16 pt
\noindent

The theory of gravitational waves is an old subject. The 
existence of general asymptotically flat radiative space-times and the 
explicit construction of various classes of solutions have received 
considerable attention over the years. They have several applications 
that serve as testing bed for the physical properties of gravitation. 
Spherical 
gravitational waves are quite interesting in this respect, as they are
thought to represent an isolated gravitationally radiating system.  
Yet, there has been no systematic 
way to understand the algebraic structure of the non-linear equations that 
govern their propagation in vacuum, apart from global considerations and 
the behavior of solutions in the asymptotic future. A few explicit solutions   
are also known to this day. It is the purpose of the present work to provide 
a new formulation of the problem by casting the corresponding sector of 
Einstein equations into zero curvature form, thus making a decisive step
toward their integration. The mathematical framework that achieves this  
purpose points to the relevance of a novel class of infinite dimensional 
Lie algebras that might be of more general value in dynamical gravitational 
problems.   

Recall that the most general gravitational vacuum solution in four 
space-time dimensions which admits a geodesic, shear-free, twist-free 
(but diverging) null congruence is described by the class of 
Robinson-Trautman metrics with line element, \cite{robi, kram},  
\be
ds^2 = 2r^2 e^{\Phi} dz d\bar{z} - 2 dtdr - H dt^2 ~, 
\ee
where $H(z, \bar{z}; t)$ has the special form
\be
H = r \partial_t \Phi - \Delta \Phi - {2m(t) \over r} - 
{\Lambda r^2 \over 3} ~.  
\ee
The affine parameter $r$ varies along the rays of the repeated null 
eigenvector and $t$ is a retarded time coordinate. Closed surfaces of 
constant $r$ and $t$ represent distorted two-dimensional spheres 
$\Sigma$ with 
metric coefficient given by $\Phi(z, \bar{z}; t)$ in the system of 
conformally flat coordinates $(z, \bar{z})$. The parameter  
$m(t)$, which in some cases represents the physical mass of the system, 
can be set equal to a constant when differs from zero; in the sequel 
$m$ is taken positive and set equal to $1/3$ by appropriate relabeling 
of the null hyper-surfaces in the Robinson-Trautman class of metrics.
The special case $m=0$ will be considered separately at the end. 
Finally, $\Lambda$ is the value of the cosmological constant in 
space-time, which can be positive, negative or zero depending on the 
physical circumstances.    

There is a close analogy between the Lienard-Wiechert fields of Maxwell 
equations and the Robinson-Trautman metrics of Einstein equations for
they both admit a principal null vector field which is geodesic, shear 
and twist free with non-vanishing divergence, thus giving a more 
intuitive meaning to the general ansatz for the metrics above, 
\cite{ted1}. In the 
gravitational case, the field equations imply that all vacuum solutions
in this class with cosmological constant $\Lambda$ satisfy the following 
parabolic fourth order differential equation for the unknown function
$\Phi$, 
\be
\partial_t \Phi = - \Delta \Delta \Phi ~, \label{main}  
\ee
where $\Delta = {\rm exp}(-\Phi)\partial \bar{\partial}$ denotes the 
Laplace-Beltrami operator on the distorted two-dimensional spheres 
$\Sigma$. Note that this non-linear equation, which is named 
after Robinson and Trautman, is not sensitive to the value of $\Lambda$.
Its time dependent solutions can be employed for the propagation of 
spherical gravitational waves in vacuum, although the strict sense 
of the word spherical symmetry will automatically imply that the metric
is static. Thus, it is more appropriate to think of such metrics as 
describing gravitational radiation outside some bounded region. 

Its fixed point solutions include the static Schwartzschild metric with  
mass $m$ (written in Eddington-Filkenstein frame) to which infinitely many
trajectories tend in the asymptotic future. Approximate solutions were first 
obtained by perturbing the Schwartzschild metric so that the geometry of the
two-dimensional slices $\Sigma$ represents distorted (e.g., axially symmetric)
configurations in multi-pole expansion. It follows, in the linearized 
approximation to equation \eqn{main}, that small perturbations associated to 
spherical harmonics with angular momentum $l \geq 2$ decay exponentially 
fast by inducing quadrapole and higher order pole radiation, and the 
configuration settles back to the round sphere as $t \rightarrow \infty$,  
\cite{ted2}. Subsequently, it was found that all dynamical solutions exhibit 
Lyapunov stability using the quadratic curvature functional 
\be
L(t) = \int_{\Sigma} R^2 d \sigma ~, \label{lyapu}  
\ee
where $R$ is the Ricci scalar curvature of $\Sigma$ and $d\sigma$ its
volume element, \cite{liap}. $L(t)$ decreases monotonically with time and the 
critical point, where it assumes its extremal value, is the round sphere
corresponding to the Schwartzschild solution. Finally, the (semi)-global 
existence and convergence of solutions with smooth initial data has 
been established in all generality, following similar arguments to the 
long time existence and convergence of solutions to the normalized 
Ricci flow, thus establishing the universal limiting character of all 
asymptotically flat Robinson-Trautman metrics, \cite{piotr}; see also 
references \cite{bern} for related earlier work. Generalizations to 
space-times with non-vanishing cosmological constant have also been 
considered in the literature, \cite{cosmo}.  
       
The Robinson-Trautman equation also arose independently in mathematics
while considering the notion of extremal metrics on compact K\"ahler 
spaces $K$, \cite{cala}.  
Using K\"ahler coordinates $(z_a, \bar{z}_a)$, the following deformation equation 
was introduced for the metric $g_{a \bar{b}}$ with Ricci scalar curvature $R$, 
\be
\partial_t g_{a\bar{b}} = \partial_a \bar{\partial}_b R ~. \label{calf}  
\ee
The quadratic curvature functional \eqn{lyapu}, which is in general defined over
the K\"ahler manifold $K$, decreases monotonically under the flow and its 
minimum defines extremal metric on it. 
When certain obstructions vanish, \cite{futa}, the 
extremal metric is canonical in that its curvature is constant.  
The equation \eqn{calf} is also known in the literature as Calabi flow and
for one-dimensional complex K\"ahler spaces, in particular, it coincides with
the main equation \eqn{main}, as first noted in reference \cite{tod}. 
In this context, the convergence of the 
Robinson-Trautman metrics to the Schwartzschild solution in the asymptotic 
future implies that the extremal metric on a two-dimensional sphere is 
the round one.     
     
Next, we sketch the algebraic framework that casts the Robinson-Trautman 
equation into zero curvature form. Let us introduce an infinite dimensional
algebra whose local part is spanned by six different elements 
$H_i(t)$, $X_i^{\pm}(t)$ with the index $i$ taking the values $0$ and $1$. 
These generators depend on a continuous real variable $t$ and assumed to  
obey the basic system of commutation relations  
\ba
& & [H_0(t), ~ X_0^{\pm}(t^{\prime})] = \pm \delta(t-t^{\prime}) 
X_0^{\pm} (t^{\prime}) ~, 
~~~~ [X_1^{\pm}(t), ~ X_0^{\mp}(t^{\prime})] = 
\delta(t-t^{\prime})H_0(t^{\prime}) ~, 
\nonumber\\
& & [H_0(t), ~ X_1^{\pm}(t^{\prime})] = \pm 
\delta(t-t^{\prime}) X_1^{\pm} (t^{\prime}) ~, 
~~~~ [X_1^+(t), ~ X_1^-(t^{\prime})] = c ~ \delta(t-t^{\prime}) H_1(t^{\prime})  
~, \nonumber\\
& & [H_1(t), ~ X_1^{\pm}(t^{\prime})] = -\partial_t \delta(t-t^{\prime}) 
X_0^{\pm}(t^{\prime}) ~,  
\ea
where $c$ is an arbitrary real constant; its physical meaning will become 
clear later. 
The remaining set of basic commutation relations are taken to be trivial, that is
impose  
\ba
& & [H_1(t), X_0^{\pm}(t^{\prime})] = 0 ~, ~~~~  
[X_0^+(t), ~ X_0^-(t^{\prime})] = 0 ~, \nonumber\\
& & [H_i(t), ~ H_j(t^{\prime})] = 0 ~.  
\ea
It can be verified that the commutation relations above are consistent
with the Jacobi identities for all values of $c$. 

The algebra can be enlarged further by taking successive 
commutators among its basic generators $X_i^{\pm}$, such as 
$[\cdots [X_i^{\pm}, X_j^{\pm}], \cdots, X_k^{\pm}]$, 
that give rise to new elements, 
as in the simpler case of Kac-Moody algebras where $X_i^{\pm}$ act as 
creation and annihilation operators in the Verma module representations; in that 
context $H_i$ are the Cartan generators of the commutative subalgebra. 
The structure of the higher commutation relations can be in principle 
established working order by order and requiring consistency with the 
Jacobi identities. 
However, there are compelling reasons to believe that the complete  
characterization of all elements of the algebra at hand, together with their
commutations relations, is a very difficult task. As will be seen later, 
its zero modes (factored by the center) generate Kac's $K_2$ 
algebra, which is the most elementary example of infinite dimensional simple
Lie algebras with infinite growth. Also, a simpler version of this 
algebra will be encountered later in a different - yet related problem - 
and it grows exponentially fast beyond the local part. Despite these 
limitations, which are also common to hyperbolic Kac-Moody algebras, 
the zero curvature formulation of the Robinson-Trautman equation can be 
established by choosing gauge connections with values only in 
the local part.    

An alternative description of the same algebraic structure is obtained by 
smearing all generators by appropriate smooth functions of compact 
support, $\varphi(t)$, and integrating over the entire range of $t$,  
\be
A(\varphi) = \int \varphi(t) A(t) dt ~,  
\ee
as commonly used in the theory of distributions and in the current algebras of 
quantum field theory. Then, the above system of 
commutation relation takes the form  
\ba
& & [H_0(\varphi), ~ X_0^{\pm}(\psi)] = \pm X_0^{\pm} (\varphi \psi) ~, 
~~~~ [X_1^{\pm}(\varphi), ~ X_0^{\mp}(\psi)] = H_0(\varphi \psi) ~, 
\nonumber\\
& & [H_0(\varphi), ~ X_1^{\pm}(\psi)] = \pm X_1^{\pm} (\varphi \psi) ~, 
~~~~ [X_1^+(\varphi), ~ X_1^-(\psi)] = c ~ H_1(\varphi \psi)  
~, \nonumber\\
& & [H_1(\varphi), ~ X_1^{\pm}(\psi)] = X_0^{\pm}(\varphi^{\prime} \psi) 
~, \label{cfcr1} 
\ea
and 
\ba
& & [H_1(\varphi), X_0^{\pm}(\psi)] = 0 ~, ~~~~  
[X_0^+(\varphi), ~ X_0^-(\psi)] = 0 ~, \nonumber\\
& & [H_i(\varphi), ~ H_j(\psi)] = 0 ~. \label{cfcr2} 
\ea
The prime here denotes derivative with respect to the continuous index $t$.

It is more convenient to use the latter form and introduce gauge connections on 
the two-dimensional space $\Sigma$ taking values in this infinite dimensional 
algebra. Postulating the zero curvature condition  
\be
F_{z \bar{z}} = [\partial + A_+ (z, \bar{z}; t), ~ \bar{\partial} + 
A_- (z, \bar{z}; t)] = 0 ~, \label{zecu} 
\ee
where the two components of the gauge connection $A_{\pm}$ 
are taken to depend on four
functions $f$, $g$, $\Phi$ and $\Psi$ of $z$, $\bar{z}$ and $t$ 
in the following way,  
\ba
A_+ & = & H_0(f) + H_1(g) + X_1^+(1) ~, \nonumber\\ 
A_- & = & X_0^- (\Psi e^{\Phi}) + X_1^- (e^{\Phi}) ~,  \label{koko} 
\ea
we arrive at a consistent system of coupled differential equations for the 
unknown functions. The computation of $F_{z\bar{z}}$ gives rise to various  
terms taking values in linear span of the generators $H_0$, $H_1$, $X_1^-$ 
and $X_0^-$ alone. These are required to vanish separately by the zero 
curvature condition \eqn{zecu} and they lead, respectively, to the four 
equations   
\ba
& & \bar{\partial} f = \Psi e^{\Phi} ~, ~~~~ \bar{\partial} g = e^{\Phi} ~, 
~~~~ \partial \Phi = f ~, \nonumber\\
& & \partial(\Psi e^{\Phi}) - f \Psi e^{\Phi} + (\partial_t g)e^{\Phi} 
= 0   \label{oucho}  
\ea
provided that the constant $c$ in the commutation relations is chosen to be 1. 

Eliminating $f = \partial \Phi$ from the first equation, it follows that
\be
\partial \bar{\partial} \Phi = \Psi e^{\Phi} ~. \label{mer1} 
\ee
Also, eliminating $f$ from the very last equation, it follows that 
$\partial_t g = - \partial \Psi$, which upon substitution into the 
$t$-derivative of second equation yields
\be
\partial_t e^{\Phi} = - \partial \bar{\partial} \Psi ~. \label{mer2} 
\ee

Note that according to equation \eqn{mer1}, the variable $\Psi$ is 
minus the Ricci scalar curvature of the two-dimensional line element
on $\Sigma$, 
\be
ds^2 = 2e^{\Phi} ~ dz d\bar{z}~ ,\label{conf}  
\ee
which has spherical topology and 
sits inside the Robinson-Trautman metric. 
When $\Psi$ is eliminated from the
combined system \eqn{mer1} and \eqn{mer2}, one readily obtains 
$\partial_t \Phi = - \Delta \Delta \Phi$, which is the same as 
\eqn{main}. Thus,  
the Robinson-Trautman equation arises as zero curvature condition  
$F_{z \bar{z}} = 0$ on the two-dimensional space of constant $(r, t)$ 
slices. The infinite dimensional algebra where the gauge connections 
assume their values is quite novel and has not been considered in the 
physics or mathematics literature before, to the best of our knowledge.
It certainly deserves further study in order to be able to construct 
explicitly conserved integrals of the equation and present its 
general solution by B\"acklund transformations. 

The situation is reminiscent, though is more complicated now, to the 
algebraic description of the Ricci flow in two dimensions using 
conformally flat (K\"ahler) coordinates. In that case, the evolution
equation $\partial_t g_{z\bar{z}} = - R_{z\bar{z}}$ for the 
$t$-dependent class of metrics \eqn{conf} reads 
$\partial_t \Phi = \Delta \Phi$. As shown in \cite{bakas1, bakas2}, 
the Ricci flow in two dimensions admits a zero curvature 
formulation, as in equation \eqn{zecu} above, choosing two-dimensional 
gauge connections of the form
\be
A_+ = H(\Psi) + X^+(1) ~, ~~~~ A_- = X^-(e^{\Phi}) ~, 
\ee
where $\partial_t \Psi = -\partial \Phi$. It was further assumed that  
the elements $H$, $X^{\pm}$ generate the local part of an infinite 
dimensional Lie algebra with basic commutation relation given (in 
smeared form) by 
\ba
& & [X^+(\varphi), ~ X^-(\psi)] = H(\varphi \psi)~, ~~~~ 
[H(\varphi), ~ H(\psi)] = 0 ~, \nonumber\\
& & [H(\varphi), ~ X^{\pm}(\psi)] = \pm X^{\pm}(\varphi^{\prime} 
\psi) ~. \label{rfcr} 
\ea
They provide the local part of a novel infinite dimensional algebra with 
anti-symmetric Cartan kernel $K(t, t^{\prime}) = \partial_t 
\delta (t-t^{\prime})$, due to the parabolic nature of the Ricci 
flow in time, whose completion by successive commutation 
of $X^+$'s and $X^-$'s yields a graded algebra with exponentially fast  
growth, \cite{savel}. 
 
The Robinson-Trautman equation or Calabi flow, as it is also known 
in the mathematics literature, is a parabolic fourth order differential 
equation that is formally related to the second order Ricci flow by 
simply taking the square root of the $t$-derivative operator. It can be 
easily seen how this procedure works and changes the order of the 
equations accordingly. Consider, in particular, the abstract 
super-evolution Ricci type flow  
\be
\partial \bar{\partial} {\cal F} = {\cal D} {\rm exp}{\cal F} ~, 
\label{sric}
\ee
where ${\cal F}(z, \bar{z}; t, \theta) = \Phi(z, \bar{z}; t) + \theta 
\Psi(z, \bar{z}; t)$ is a mixed superfield with bosonic components 
$\Phi$ and $\Psi$ and ${\cal D} = \partial / \partial \theta - \theta 
\partial / \partial t$ is the superderivative in $(1,1)$ superspace 
with coordinates $(t, \theta)$, so that $\theta^2 = 0$ and 
${\cal D}^2 = -\partial / \partial t$. Expanding ${\rm exp}{\cal F} = 
(1 + \theta \Psi){\rm exp} \Phi$ and comparing the even and odd parts 
of the super-flow \eqn{sric}, one immediately arrives at the system
\eqn{mer1} and \eqn{mer2} that amount to the Robinson-Trautman 
equation for the variable $\Phi$. 
Note on the side that the procedure of taking the 
square root of the time derivative operator might be of more general 
value relating other pairs of second and fourth order parabolic equations.  

According to this observation, it is natural to expect that there is 
an infinite dimensional superalgebra at work, which generalizes the  
commutation relations
\eqn{rfcr} and accounts for the (seemingly ad hoc) choice of the bosonic 
algebra \eqn{cfcr1} and \eqn{cfcr2} used in the theory of spherical 
gravitational waves. By
the same token, it is also natural to expect that the general 
solution can be obtained in closed form using group theoretical methods
similar to those employed for the Ricci flow, \cite{bakas1, bakas2}. 
It should be emphasized
in both cases that the parabolic nature of the equations, which tend to
dissipate curvature perturbations from the canonical form of the metric, 
is not in contradiction with their integrability. The latter refers to their
zero curvature description in two dimensional spaces with local 
coordinates $(z, \bar{z})$, whereas the deformation variable $t$ is treated
differently by accommodating it into the structure of the Lie algebra where
the gauge connections take values. 
The reciprocity of our description is common to other examples  
of non-linear differential equations. Note, however, that the underlying 
algebras do not necessarily possess a non-zero invariant bilinear form, 
in contrast to the case of Kac-Moody algebras with symmetrizable Cartan 
matrices. Thus, the exact mathematical meaning of integrability 
and the construction of conserved integrals in two dimensions  
present some challenging questions. Further details on these 
issues will be given elsewhere, \cite{bakas3}.

We turn now to the special case of Robinson-Trautman metrics with mass 
parameter $m=0$ for which the main equation \eqn{main} simplifies to
\be
\Delta \Delta \Phi = 0 ~. \label{zema}  
\ee
Equivalently, it can be stated as 
$\Delta \Phi (z, \bar{z}; t) = \chi(z, t) + \bar{\chi}(\bar{z}; t)$,  
where $\chi$ and $\bar{\chi}$ are holomorphic and anti-holomorphic 
functions, respectively, that are complex conjugate to each other. 
Note here that $t$ appears as spectator with no apparent
role in the dynamics. This class of solutions is naturally related to 
Nirenberg's mathematical problem for finding metrics of prescribed 
curvature on $S^2$, since $R[S^2] = -\Delta \Phi$; see, for instance, 
\cite{stru}, for a recent approach based on flows. The zero curvature 
formulation of equation \eqn{zema} is slightly different as it requires
the choice of gauge connections \eqn{koko}  
taking values in the infinite dimensional algebra \eqn{cfcr1}, 
\eqn{cfcr2} with parameter $c=0$. This, in turn, implies that the resulting 
system is the same as in equation \eqn{oucho} given above, apart from 
the component $\bar{\partial} g = {\rm exp} \Phi$ that is now replaced 
by $\bar{\partial} g = 0$. 
Eliminating $f$ and $g$, as before, it follows 
\be
\partial \bar{\partial} \Phi = \Psi e^{\Phi} ~, ~~~~ \partial \bar{\partial} 
\Psi = 0 
\ee
that yield equation \eqn{zema} as required. Finally, note that the  
one-parameter family of holomorphic 
functions $g$ and $\chi$ are simply related to each other by 
$\partial \chi = - \partial_t g$.   

In conclusion, it is instructive to compare our results with other works
on the integration of the Robinson-Trautman equation 
by prolongation methods. Type II 
solutions (in terms of Petrov classification), 
which are the generic examples when 
the mass parameter $m \neq 0$, have resisted all previous 
attempts with the exception of some partial (yet inconclusive) results 
described in reference \cite{type}. 
On the other hand, type III solutions, which have $m=0$, have been more 
tractable by such methods.   
The corresponding metrics were shown to exhibit 
a rich prolongation structure given by the simple infinite dimensional 
algebra $K_2$ (in Kac's notation), \cite{finl}. This result fits nicely 
into the algebra \eqn{cfcr1}, \eqn{cfcr2} with $c=0$, since the zero modes  
of the currents $H_i(t)$, $X_i^{\pm}(t)$ form the local part of $K_2$ 
algebra. Introducing the change of basis
\ba
& & e_0 = {1 \over \sqrt{2}} (X_0^+ + X_1^+) ~, ~~~~ e_1 = {1 \over \sqrt{2}} 
(-X_0^+ + X_1^+) ~, \nonumber\\
& & f_0 = {1 \over \sqrt{2}} (X_0^- - X_1^-) ~, ~~~~ f_1 = {1 \over \sqrt{2}} 
(X_0^- + X_1^-) ~, 
\ea
it easily follows that the zero modes of the currents, without $t$-dependence,
satisfy the basic system of commutation relations
\be
[e_i, ~ f_j] = \delta_{ij} h ~, ~~~~ [h, ~ e_i] = e_i ~, ~~~~ 
[h, ~ f_i] = - f_i 
\ee
with $h$ given by the zero mode of $H_0$. These are precisely 
the defining relations for the local part
of the algebra $K_2$, \cite{kac}, whereas the zero mode of the generator $H_1$ 
commutes with the rest and decouples. 

For type II metrics with mass parameter $m$, the commutation relations 
\eqn{cfcr1}, \eqn{cfcr2} have $c=3m$, in which case the zero mode of the current 
$H_1$ generates the center of the algebra with coefficient 
(central charge) $c$. Thus, $K_2$ also  
describes the algebra of the zero modes of the currents in type II situations  
provided that the one-dimensional center is factored out.  
In either case, the infinite dimensional algebra we introduced in this paper 
is much larger, as it includes an infinite number of Fourier modes, 
and (apparently)  
provides the minimal choice for establishing the zero curvature formulation 
of the Robinson-Trautman equation for spherical gravitational waves in vacuum 
for all values of the mass parameter and the cosmological constant.

It will be interesting to see how the results may generalize to 
Einstein-Maxwell theory. Particular interest also present the generalizations 
in the presence of additional massless fields that arise in string theory, 
\cite{guve}, and their possible implications for the gravity/gauge correspondence
beyond the plane wave limit. 

\vskip .5in
\centerline{\bf Acknowledgments}

This work was supported in part by the European Research and Training Network
``Constituents, Fundamental Forces and Symmetries of the Universe" 
under contract number MRTN-CT-2004-005104 and the INTAS program ``Strings, 
Branes and Higher Spin Fields" under contract number 03-51-6346. I thank 
Ecole Polytechnique for hospitality in the initial stages of the present
work and CNRS for financial support. I also thank
the Theory Division at CERN for hospitality and financial support during 
my sabbatical leave in the academic year 2004-05, where the main body of 
the present work was carried out in excellent and stimulating enviroment.

\end{document}